\documentclass[aps,prd,twocolumn,reprint,superscriptaddress,longbibliography,nofootinbib,floatfix]{revtex4-1}
\usepackage{epsfig}
\usepackage{amsmath,amssymb,amsfonts}
\usepackage{hyperref}
\usepackage{mathrsfs}
\usepackage{bbm}
\usepackage{slashed}
\usepackage{graphicx}
\usepackage{verbatim}
\usepackage[usenames]{color}

\newcommand{\be}{\begin{equation}}
\newcommand{\ee}{\end{equation}}
\newcommand{\bw}{\begin{widetext}}
\newcommand{\ew}{\end{widetext}}
\newcommand{\bi}{\begin{itemize}}
\newcommand{\ei}{\end{itemize}}
\newcommand{\bea}{\begin{eqnarray}}
\newcommand{\eea}{\end{eqnarray}}
\newcommand{\ud}{\mathrm{d}}

\newcommand{\LCm}{{\scriptscriptstyle -}} 
\newcommand{\LCp}{{\scriptscriptstyle +}}
\newcommand{\LCpm}{{\scriptscriptstyle \pm}}

\newcommand{\LCperp}{{\scriptscriptstyle \perp}}

\newcommand{\A}{\mathcal{A}}
\newcommand{\snitt}[1]{\langle #1 \rangle}

\usepackage[T1]{fontenc} \usepackage[latin1]{inputenc}

\begin{document}
\title{Pair production from residues of complex worldline instantons}
\author{Anton Ilderton}
\email[]{anton.ilderton@chalmers.se}
\affiliation{Dept.~Applied Physics, Chalmers University of Technology, SE-41296 Gothenburg, Sweden}

\author{Greger Torgrimsson}
\email[]{greger.torgrimsson@chalmers.se}
\affiliation{Dept.~Applied Physics, Chalmers University of Technology, SE-41296 Gothenburg, Sweden}

\author{Jonatan Wårdh}
\email[]{jontan@student.chalmers.se}
\affiliation{Dept.~Applied Physics, Chalmers University of Technology, SE-41296 Gothenburg, Sweden}

\begin{abstract}
We study nonperturbative pair production in electric fields with lightlike inhomogeneities, using complex worldline instantons. We show that the instanton contribution to the pair production probability is a complex contour integral over the instanton itself, and that pair production in the considered fields can be recast in terms of Cauchy's residue theorem. The instantons contribute residues from the poles they circulate (i.e.~give local contributions), and the invariance of complex integrals under contour deformation manifests in the instanton contributions as invariance under a set of generalised, complex, reparameterisations.
\end{abstract}
\maketitle
\section{Introduction}
There is a drive to better understand pair production in electromagnetic fields, spurred both by its nonperturbative nature~\cite{Sauter:1931zz,Heisenberg:1935qt,Schwinger:1951nm} and by experimental prospects for observing pair production using intense lasers~\cite{Dunne:2008kc,Heinzl:2008an,Bulanov:2010ei,DiPiazza:2011tq,Adorno:2014bsa,Gonoskov:2013ada,Kohlfurst:2013ura,Blinne:2013via,Strobel:2014tha}.

Worldline path integral methods have proven powerful for such investigations~\cite{Affleck:1981bma,Gies:2003cv,Dunne:2005sx,Dunne:2006st,Gordon:2014aba} (and have a wealth of other applications~\cite{Schubert:1996jj,Dietrich:2013kza,Mansfield:2014vea,Edwards:2014bfa}). There are though few exact analytic results for pair production in realistic fields with multi-dimensional inhomogeneities~\cite{Dunne:2006ur,Schneider:2014mla}. In order to better understand this difficult problem, it seems sensible to exhaust our knowledge of the three simplest cases, namely fields depending on a single timelike, spacelike, or lightlike coordinate. 

In the lightlike case the pair production probability, or rather the imaginary part of the effective action, is given exactly by a locally constant approximation~\cite{Tomaras:2000ag,Tomaras:2001vs}. This surprising simple result does not extend to time-dependent or position-dependent fields~\cite{Naroz,Dunne:2004nc}, but has been rederived using both functional~\cite{Fried:2001ur} and worldline methods~\cite{Ilderton:2014mla}. Our goal is therefore not to give another derivation, but to understand more about why localisation occurs. The simplicity of the result suggests that a symmetry may be at play; this is an intriguing prospect given the fundamental importance of symmetry in QFT.

We will uncover this symmetry below, by considering pair production in the language of worldline instantons. These are periodic solutions to the classical equations of motion, and their classical action gives the dominant, nonperturbative, contribution to the effective action~\cite{Affleck:1981bma,Gies:2003cv,Dunne:2005sx,Dunne:2006st}. Worldline instantons are typically taken to be \textit{real} loops in {\it Euclidean} space but, as described in detail in~\cite{Dumlu:2011cc}, they will in general be complex.  (Complex instantons are widely studied in more general contexts, see~\cite{Kim:2007pm,Basar:2013eka,Tanizaki:2014xba,Cherman:2014sba} and references therein.) We will see that fields with lightlike inhomogeneities offer an ideal system for studying complex worldline instantons, and by doing so we will be able to reveal new structure.

We will show that the contribution of these complex instantons to the effective action are contour integrals over the instantons themselves, and are residues from poles, i.e.~points. Further, the instanton contributions are invariant under what can be viewed as a complex extension of the reparameterisation invariance which underlies the worldline formalism~\cite{Polyakov:1987ez,Mansfield:1990tu}. We will show that this symmetry, which simply corresponds to the freedom to deform integration contours in the complex plane as per the residue theorem, is responsible for the localisation of the instanton contributions to the classical action.

This paper is organised as follows. In Sect.~\ref{Background} we briefly review some necessary background and describe our field model. We then present the complex instanton solutions for constant electric fields and show that their contribution to pair production admits a boost-like symmetry. In Sect.~\ref{IRES} we reveal the complex structure in the instantons of arbitrary lightfront-time dependent fields, present explicit solutions for the Sauter and oscillating fields, and discuss their generalised reparameterisaton invariance. We conclude in Sect.~\ref{Conc}.
\section{Background}\label{Background}
The pair production probability in an electromagnetic field is in the worldline approach built from instantons. These are periodic solutions to the equations of motion of the classical worldline action $S$,
\be\label{S-original}
	S = -\frac{m^2}{2} T -\int\limits_0^1\!\ud\tau\, \frac{\dot{x}.\dot{x}}{2T} +e \int\limits_0^1\!\ud\tau\ \dot{x}. A(x) \;,
\ee
where $\tau$~parameterises the worldline, a dot is a $\tau$-derivative, and $T$ is proper time.

Define lightfront coordinates $x^\LCpm = t\pm z$, $x^\LCperp=\{x,y\}$, then our electric field is $E(x^\LCp) = -2\partial_\LCp A_\LCm(x^\LCp)$, polarised in the $z$--direction.  Two motivations for studying these fields are, as already stated, understanding the localisation of the effective action which is seemingly particular to the lightlike case, and the study of complex rather than (Euclidean--) real instantons. The instantons in the lightlike case will necessarily be complex because rotating $E(x^\LCp) = E(t+z)$ to Euclidean space would introduce both real and imaginary parts into the action (\ref{S-original}). For this reason nothing is gained by explicitly introducing Euclidean variables, and we therefore work with Minkowski variables throughout.

As phenomenological motivation one may argue as follows: consider two colliding, transverse, laser pulses travelling in the $\pm z$ directions, i.e.~depending essentially on $t\pm z$. It is common to take a time-dependent field $E(t)$ as a rough model for the standing wave formed in the focus of these two colliding pulses. Now consider instead a single laser pulse, depending on $t+z$, which is \textit{focussed}. Focussing introduces a {\it longitudinal} electric field which will also depend on $t+z$, as well as acquiring e.g.~a $z$-dependence describing the focussing~\cite{Davis}. It is mainly this component of the field which can be responsible for pair production. We therefore take the longitudinal field $E(x^\LCp)$ to be a rough model of the longitudinal field in a focussed laser pulse. Note that neither our fields nor time-dependent electric fields $E(t)$ obey Maxwell's equations in vacuum. This is unavoidable if one is to make analytic progress, however. For a recent discussion of this issue see~\cite{Linder:2015vta}.

Returning to (\ref{S-original}), we wish to solve the equations of motion. To keep this discussion concise we eliminate $T$ using a saddle point approximation, following~\cite{Affleck:1981bma,Dunne:2005sx}, which gives a nonlocal action. Periodicity requires that the $x^\LCperp$ are constant here, and they decouple.  Writing $eE/m \equiv \mathcal{E}$, the only nontrivial equations of motion are then
\be\label{XEOM-ickelokal}
	\ddot{x}^\LCpm = \pm i a \mathcal{E}(x^\LCp)\dot{x}^\LCpm \;,\qquad \dot{x}^2 = \dot{x}^\LCp \dot{x}^\LCm = -a^2 \;.
\ee
Periodic solutions to (\ref{XEOM-ickelokal}) exist only for certain $a$. We will look for solutions with $a$ real and positive because then the classical action and pair production probability $\mathbb{P}$ evaluated on the solution become, as in~\cite{Dunne:2005sx},
\be\label{SA}
	i S = -\frac{ma}{2} \quad\text{ and } \quad \mathbb{P} \sim \exp \bigg(-\frac{ma}{2} \bigg)\;.
\ee
\subsection{Constant fields}
\begin{figure}[t!]
\includegraphics[width=0.44\columnwidth]{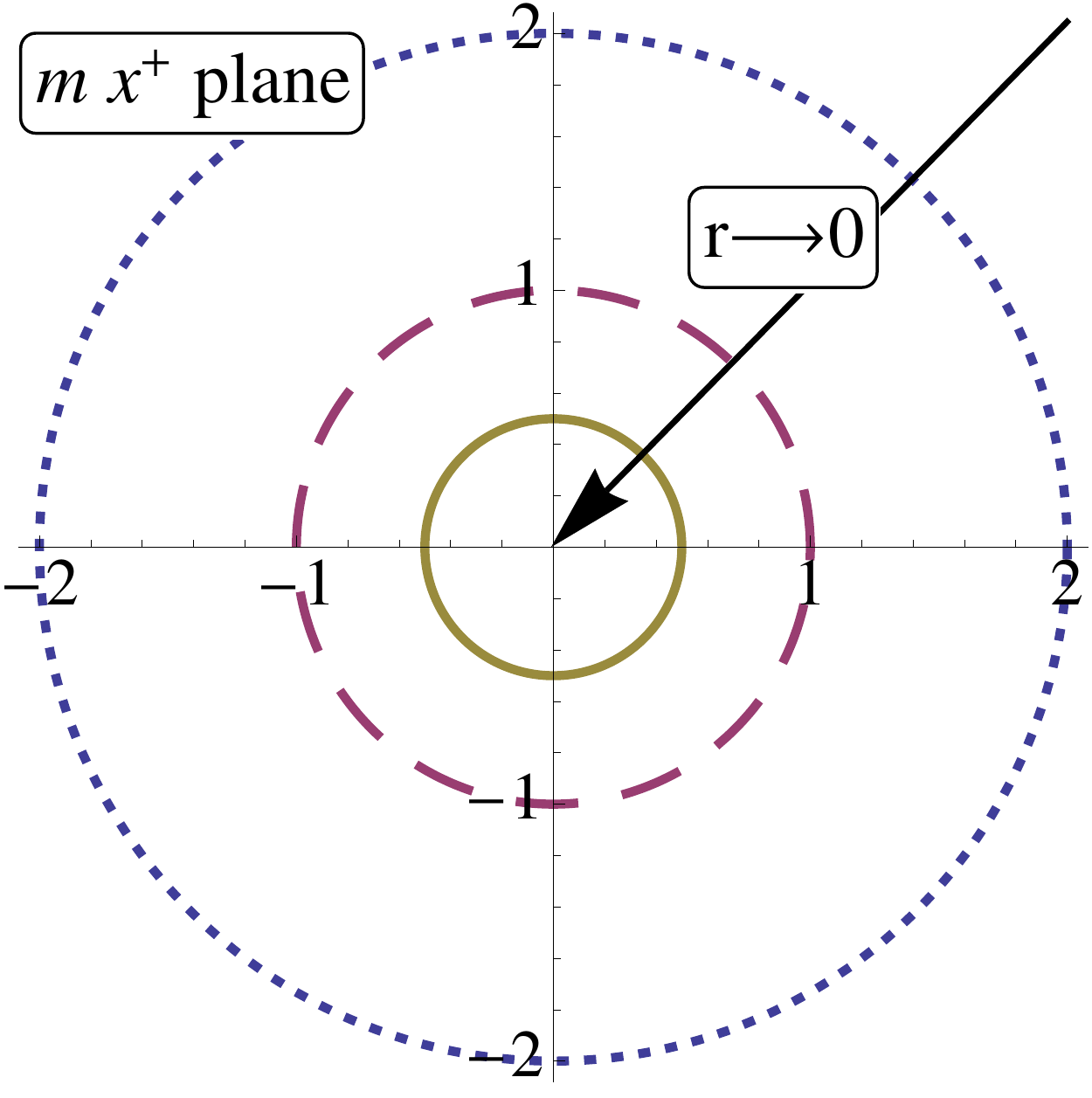}
\includegraphics[width=0.44\columnwidth]{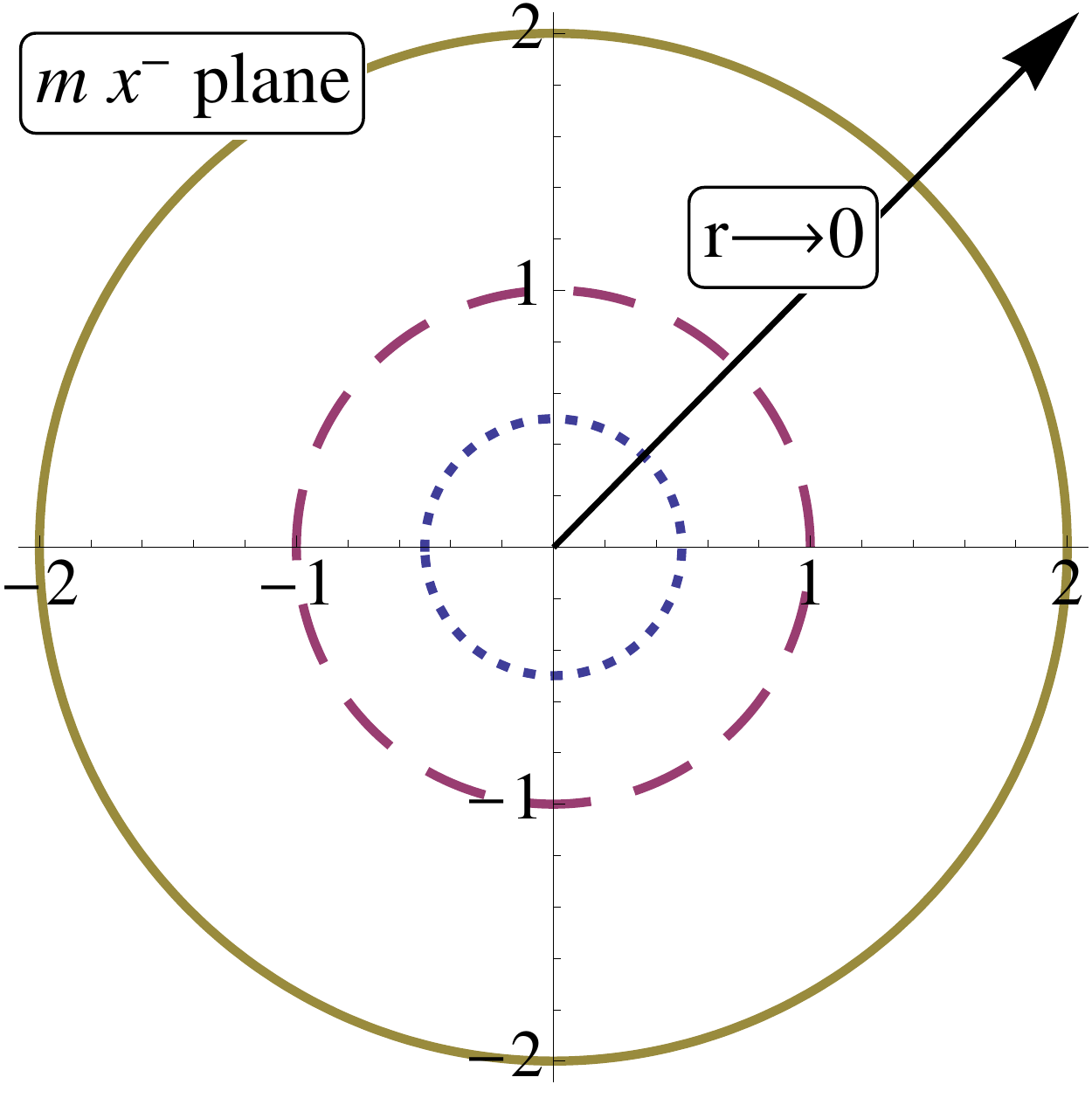}
\caption{\label{FIG:KONST} The complex instantons in a constant electric field for $r=2, 1,1/2$ (dotted, dashed, solid) and $\mathcal{E}=m$. As $r\to 0$  the $x^\LCp$ instanton shrinks to a point, while the $x^\LCm$ instanton expands.}
\end{figure}
Consider first the instanton solutions to (\ref{XEOM-ickelokal}) for constant fields. The real solutions are circles in (Euclidean) $t$--$z$ space of fixed radius $1/\mathcal{E}$~\cite{Affleck:1981bma,Dunne:2005sx}. As observed in~\cite{Dumlu:2011cc} though, instantons will in general be complex. By studying these general solutions we will be able to reveal new structure. For constant fields the general periodic solution to (\ref{XEOM-ickelokal}) is
\be\label{SOL:KONST}
	x^\LCp(\tau) = c^\LCp +  \frac{1}{m}r e^{2n\pi i \tau},\ x^\LCm(\tau) = c^\LCm  -\frac{m}{\mathcal{E}^2r} e^{-2n\pi i \tau} \;,
\ee
where periodicity requires $a=2n\pi/\mathcal{E}$, and dimensionless~$r$ can be taken real and positive, as any phase in $r$ can be absorbed into a $\tau$--reparameterisation. The instantons (\ref{SOL:KONST}) are complex, being circles in both the complex $x^\LCp$ and $x^\LCm$ planes. Their centres $c^\LCpm$ are arbitrary as this system is translation invariant. The radii of the circles are inversely proportional, but also arbitrary, see Fig.~\ref{FIG:KONST}. (To recover the solution in~\cite{Affleck:1981bma,Dunne:2005sx} take $r=m/\mathcal{E}$ and rotate $t\to it$.) The classical action (\ref{SA}) is independent of $r$, as is the mass-shell relation in (\ref{XEOM-ickelokal}). Note that changing $r\to r'$ is equivalent to rescaling, for $\varphi:=\log (r'/r)$, 
\be\label{boost1}
	\dot{x}^\LCp \to e^\varphi \dot{x}^\LCp \;,\qquad   \dot{x}^\LCm \to e^{-\varphi} \dot{x}^\LCm \;,
\ee
which has the same form as a Lorentz boost of the instanton momenta (in lightfront coordinates), rapidity $\varphi$ in the $z$-direction. The Lorentz invariant $\dot{x}^2$ and the classical action are naturally boost invariant. We explain below what lies behind this invariance, and how it extends to inhomogeneous electric fields depending on~$x^\LCp$.
\section{Instantons and residues}\label{IRES}
To begin, define $\langle \ldots \rangle$ to be the proper-time average over $\tau\in[0,1]$.
Integrating the $x^\LCp$--equation of motion in (\ref{XEOM-ickelokal}) with an integrating factor gives an implicit relation from which $a$ and the pair production probability (\ref{SA}) are determined by periodicity:
\be\label{PERIODICITET}
	\dot{x}^\LCp(1) = e^{i a \langle \mathcal{E} \rangle}\dot{x}^\LCp(0) \implies  a = \frac{2n\pi}{\langle \mathcal{E} \rangle} \;,
\ee
where $\text{sign}(n)=\text{sign}(\langle\mathcal{E}\rangle)$ so that $a>0$. Here $n\in\mathbb{Z}$ is the winding number of the velocity $\dot{x}^\LCp$ about the origin, as seen by writing $2n\pi i = \langle i a \mathcal{E} \rangle =\langle \ddot{x}^\LCp / \dot{x}^\LCp\rangle$. $n$ is therefore the `turning number' of the instanton $x^\LCp$~\cite{DG}. Integrating (\ref{XEOM-ickelokal}) directly gives
\be\label{A}
	\dot{x}^\LCp(\tau) = i a \mathcal{A}(x^\LCp(\tau)) \;,
\ee
in which the potential $\mathcal{A}$ obeys $\mathcal{A}' = \mathcal{E}$. Note that there is no (gauge) freedom in the integration constant: integrating (\ref{A}), the requirement that $x^\LCp$ be periodic fixes the integration constant such that $\langle\mathcal{A}\rangle=0$; the constant is therefore dependent on the instanton solution itself. See also~\cite{Dunne:2005sx,Dunne:2006st}. (It follows from~(\ref{A}) that the instantons will not self-intersect.)

Our key observation is that all expectation values can be rewritten as contour integrals in the complex plane. Consider for example the identity
\be\label{1}
	1 = \langle 1 \rangle = \int\limits_0^1\!\ud\tau\ \frac{\dot{x}^\LCp}{\dot{x}^\LCp} = \frac{n}{ia} \oint\limits_\text{inst.}\frac{\ud z}{\A(z)} \;,
\ee
where the contour is the instanton itself. $n$ appears here as the number of times the parameterisation covers the closed curve; for simple curves, this is indeed equal to the turning number~\cite[\S5.7]{DG}. It follows from (\ref{1}) that there must be at least one pole $z=z_*$ within the instanton loop at which the potential $\mathcal{A}(z_*) = 0$. We assume for simplicity that $\mathcal{A}(z)$ is analytic within the loop, and that $1/\mathcal{A}(z)$ has a single simple pole there: this will be the case for our later examples. Extensions will be discussed below and considered in~\cite{Ilderton:2015qda}.

Combining (\ref{1}) and (\ref{PERIODICITET}) shows that the classical action of an instanton is the residue from the pole at $z_*$:
\be\label{RES1}
	\frac{1}{\langle \mathcal{E}\rangle} = \frac{1}{2\pi i} \oint\limits_\text{inst.}\frac{\ud z}{\A(z)} = \frac{1}{\mathcal{E}(z_*)} \;.
\ee
\begin{figure*}[t!!]
\centering\includegraphics[width=0.51\columnwidth]{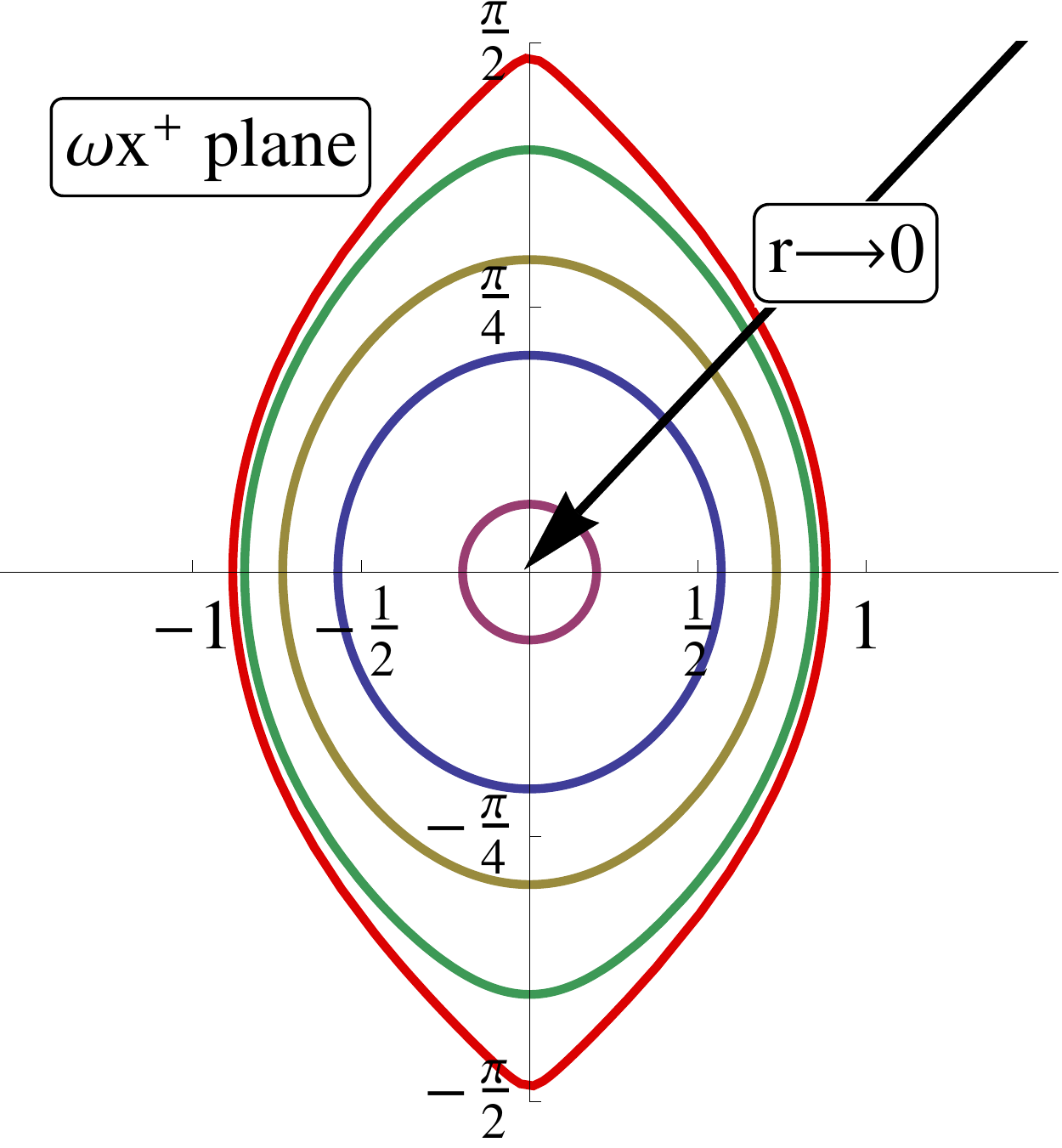}
\includegraphics[width=0.54\columnwidth]{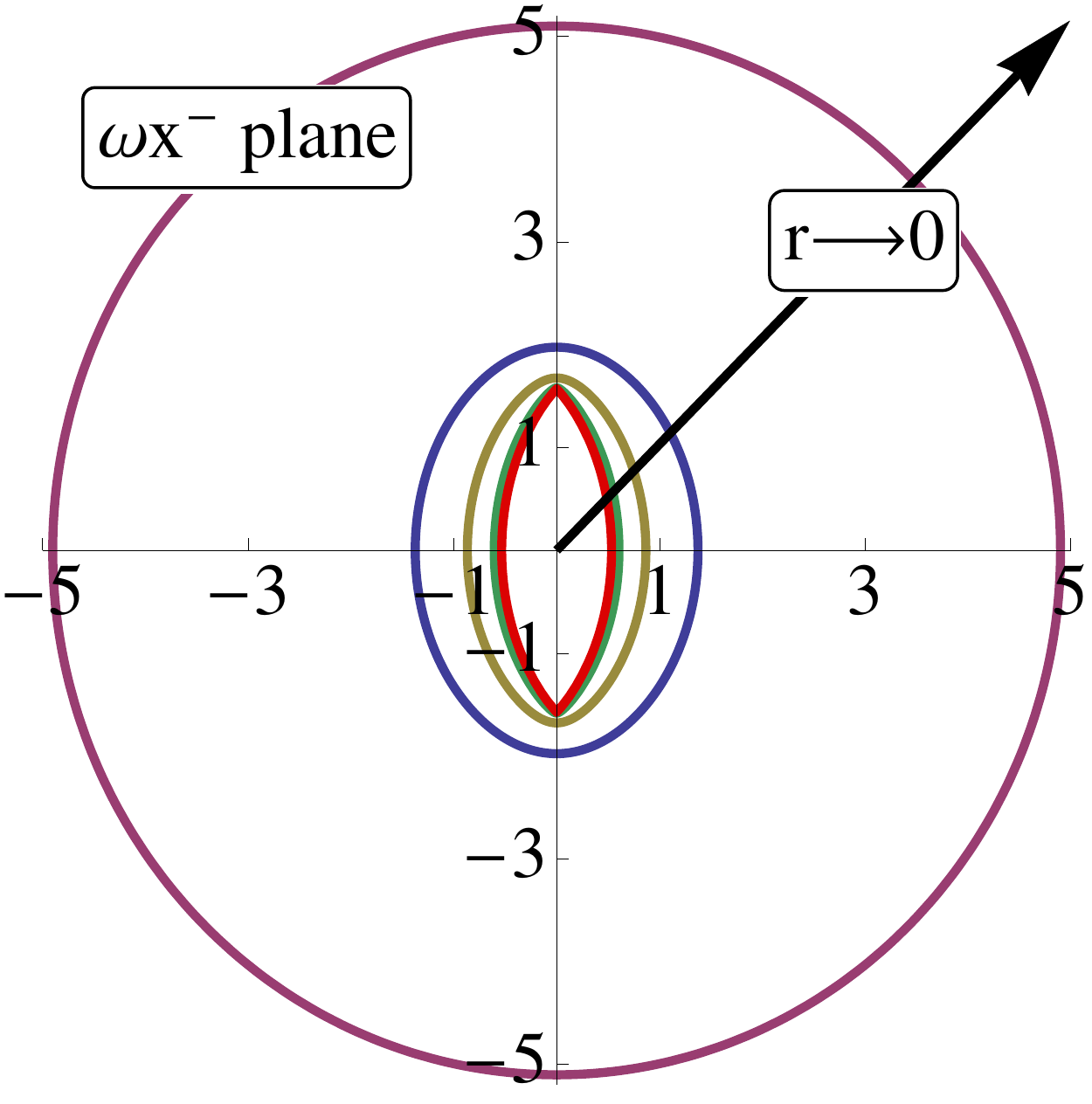}
\includegraphics[width=0.64\columnwidth]{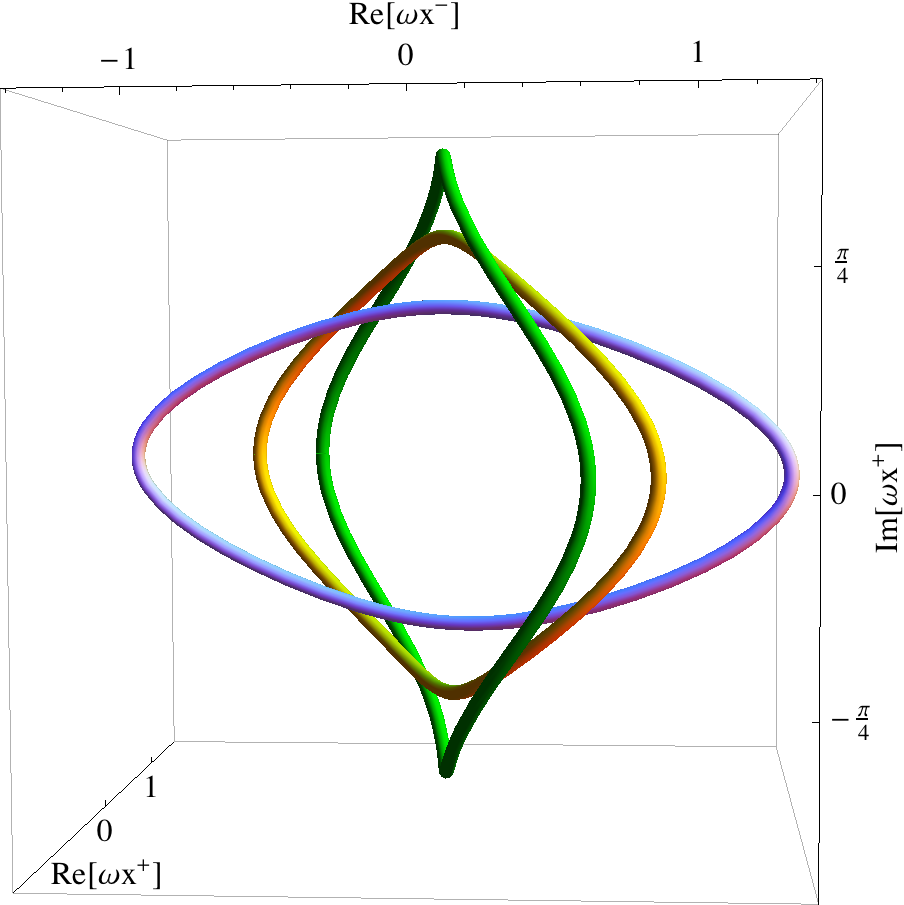}
\caption{\label{FIG:SECH1}  \textit{Left}: The $x^\LCp$ instantons in the Sauter pulse for $r=1/5,3/5,4/5,19/20,999/1000$. As $r\to 0$ the instantons contract to the origin, while as $r\to 1$ they are reflected from the poles of the electric field at $\omega x^\LCp = \pm\pi/2$~\cite{Schneider:2014mla}. \textit{Middle}: The $x^\LCm$ instantons for the same values of $r$ and $\omega=\mathcal{E}$, which expand as the $x^\LCp$ instantons contract. \textit{Right}: A projection of the instantons in $x^\LCp$--$x^\LCm$ space for $r=3/5, 4/5, 19/20$ (blue, yellow, green).}
\end{figure*}

The instanton contribution to the pair production probability therefore comes from a \textit{point} in the complex $x^\LCp$ plane, even though the instanton itself is a nontrivial loop. The pole is located at the centroid of the instanton, $\langle x^\LCp \rangle$:
\be\label{RES2}
	\langle x^\LCp \rangle = \frac{n}{ia} \oint\limits_\text{inst.}\!\ud z\ \frac{z}{\A(z)} = \frac{\snitt{\mathcal{E}}}{\mathcal{E}(z_*)} z_* = z_* \;.
\ee
Combining (\ref{RES1}) and (\ref{RES2}) gives
\be
	\langle \mathcal{E}(x^\LCp)\rangle = \mathcal{E}(\snitt{x^\LCp}) \;.
\ee
This tells us that the instanton contributions to the effective action {\it localise}. This happens because the instantons circulate poles in the complex plane, and therefore contribute only residues from points. Thus we see why the nonperturbative exponential part of the pair production probability localises in lightfront-time dependent fields. In order to extend this result to the whole effective action we would clearly need to see how localisation occurs in the `prefactor' contributions from fluctuations around the instantons. However, as stated earlier, we are interested here in understanding more about how and why localisation arises, rather than recovering known final results. We therefore  continue to focus on the structure and symmetries of the instantons themselves.

We turn now to the $x^\LCm$ solution. Integrating the mass-shell condition $\dot{x}^\LCm = -a^2/\dot{x}^\LCp$, periodicity requires
\be
	 0 = x^\LCm(1) - x^\LCm(0) =  \oint\limits_\text{inst.}\!\frac{\ud z}{\A^2(z)}  \;.
\ee
Picking up the pole at $\langle x^\LCp \rangle$ implies we must have $\mathcal{E}'(\langle x^\LCp\rangle) = 0$. The instantons therefore circulate the electric field extrema. This is a consequence of using the saddle-point approximation for $T$ -- for extensions see~\cite{US2}.

We can confirm all this structure for constant $\mathcal{E}$ by using (\ref{SOL:KONST}), as all $\tau$-integrals can be performed explicitly. We have $\snitt{x^\LCp} = c^\LCp$, so the potential obeying $\snitt{\mathcal{A}}=0$ is
\be
	\mathcal{A}(z) = \mathcal{E} (z - c^\LCp) \;.
\ee
The inverse potential has a simple pole at $c^\LCp = \langle x^\LCp \rangle$ with residue $1/\mathcal{E}$. Note the importance of the constant term in the potential.  As $1/\mathcal{A}^2$ has no residue, $\mathcal{E}'\equiv 0$ here, $x^\LCm$ is periodic for arbitrary instanton positions, consistent with~(\ref{SOL:KONST}).

We have now seen that the instanton contribution to pair production is a contour integral over the complex instanton itself.  By the residue theorem, though, the value of the integral is invariant under deformation of the contour. In particular, it is invariant as we contract the contour around the pole. Remarkably, this freedom is found in the instantons themselves. For a constant field, variation of $r$ describes that subset of deformations for which the contour remains an (instanton) solution of the equations of motion. Eq.~(\ref{SOL:KONST}) shows that reducing $r$ means uniformly contracting $x^\LCp$ to its centroid, i.e.~contracting the contour around the pole of the integrand in~(\ref{RES1}). Thus the invariance of the classical action under changes in~$r$ is simply the well-known statement that the value of a contour integral is independent of the exact form of the contour. We now verify these arguments for inhomogeneous fields using explicit examples.
\subsection{Sauter pulse}
Our first example is the Sauter pulse,
\be
	\mathcal{E}(x^\LCp) = \mathcal{E}_0\, \text{sech}^2 \omega x^\LCp\;.
\ee
 The instanton solutions are, for $\mathcal{E}_0>0$ and $n\in\mathbb{Z}^\LCp$, 
\be\label{SOL:SAUTER}
	\omega x^\LCp = \sinh^{-1} r e^{2n\pi i \tau} \;, \quad a = {2n\pi}/{\mathcal{E}_0} \;.
\ee
$x^\LCm$ is found by integrating $-a^2/\dot{x}^\LCp$, but the explicit form is unrevealing. For small $r$ the instanton (\ref{SOL:SAUTER}) is approximately circular, as for a constant field.  As $r\to 1$ from below, the instanton becomes elliptic and then `sharp', see Fig.~\ref{FIG:SECH1}. As observed in~\cite{Schneider:2014mla}, the instanton is being reflected from the pole of the electric field at $\omega x^\LCp = \pm i\pi/2$. If we try to push the instantons past this pole by taking $r=1$ then the argument in (\ref{SOL:SAUTER}) crosses the branch points of $\sinh^{-1}$ on the imaginary axis, and fails to be periodic. This describes the `breaking' of the instanton across the singularity in the field.

A direct calculation shows that the centroid $\langle x^\LCp\rangle$ is independent of $r$. Indeed $\langle x^\LCp\rangle =0$, to which the instanton contracts as $r\to 0$. We can show explicitly that the action localises on the pole contribution:
\be\label{bob}
	\langle \mathcal{E} \rangle = \mathcal{E}_0\bigg[ \tau+i\frac{\log [1+r^2 e^{4 i \pi  n \tau}]}{4 \pi  n}\bigg]\bigg|_0^1 = \mathcal{E}_0 = \mathcal{E}(\langle x^\LCp \rangle) \;.
\ee
Under changes in $r$, the instanton velocities $\dot{x}^\LCpm$ do not transform as simply as in (\ref{boost1}) -- we return to this below, once we have seen a second example. We have however the same invariances as for the constant field case: both the Lorentz scalar $\dot{x}^2$ and the classical action are $r$--independent, see (\ref{bob}). Taking $r\to 0$ again contracts the integration contour in (\ref{RES1}) around the pole of the integrand, through instanton solutions to the equations of motion.

\subsection{Sinusoidal field}
\begin{figure}[t!]
\includegraphics[width=0.7\columnwidth]{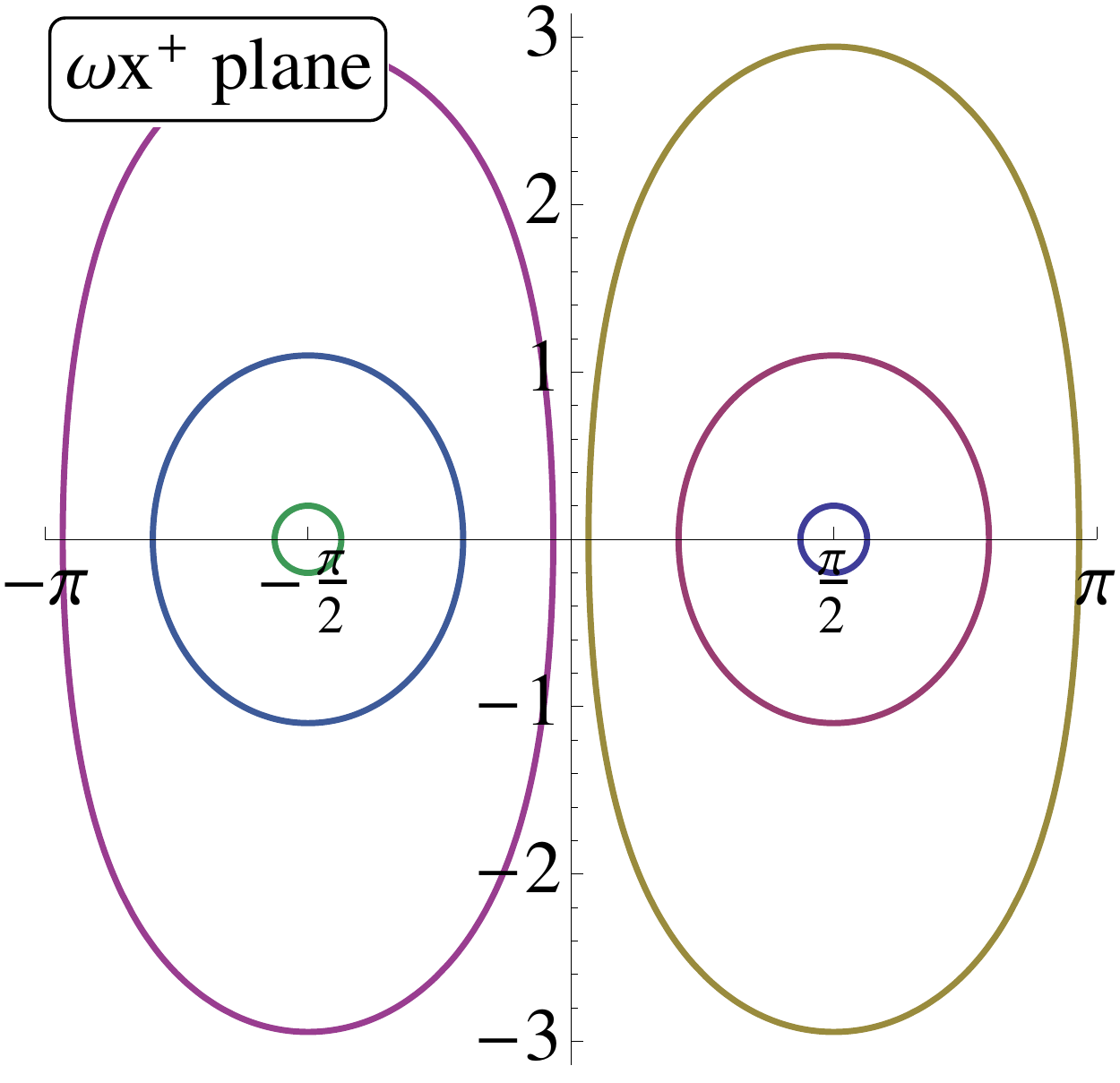} \\
\vspace{20pt}
\includegraphics[width=0.7\columnwidth]{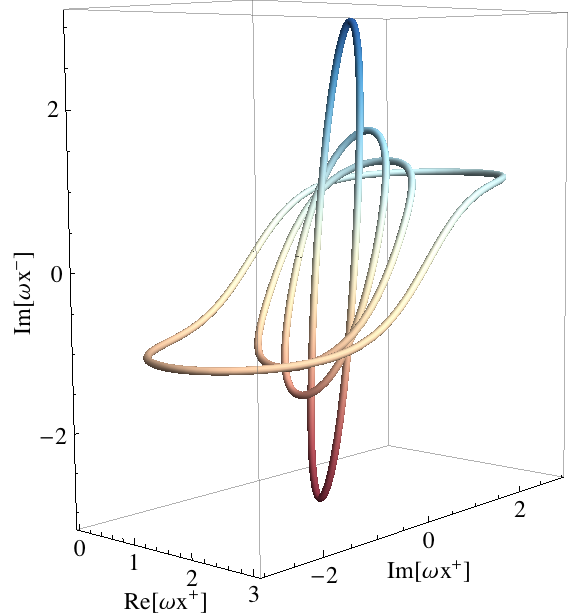}
\caption{\label{FIG:SINUS1} \textit{Upper panel:} The two instantons (\ref{SOL:SINUS}) in the sinusoidal field (\ref{E:SINUS}), for $r=1/10,1/2,99/100$ (expanding outward). The instantons circulate the electric field maxima at $\omega x^\LCp= \pm \pi/2$. \textit{Lower panel:} A projection of the instantons in $x^\LCp$--$x^\LCm$ space, circulating $\omega x^\LCp= \pi/2$ and $x^\LCm = 0$, for $\omega=\mathcal{E}$ and various $r$.}
\end{figure}

The original derivation of the effective action in the considered system was only for electric fields of fixed sign~\cite{Tomaras:2000ag,Tomaras:2001vs}. One might imagine that locality is lost if one abandons this assumption.  We therefore consider here an exactly soluble case where the field indeed changes sign. We consider the sinusoidal field
\be\label{E:SINUS}
	\mathcal{E}(x^\LCp) = \mathcal{E}_0 \sin \omega x^\LCp \;,
\ee
which has two extrema per cycle. Taking $\mathcal{E}_0>0$ the two corresponding instantons are, with $n\in\mathbb{Z}^\LCp$, 
\be\label{SOL:SINUS}
	\omega x^\LCp(\tau) = 2 \tan^{-1}\pm\frac{1-r e^{\pm 2n\pi i \tau} }{1+r e^{\pm 2n\pi i \tau}} \;, \quad a = 2n \pi/\mathcal{E}_0 \;.
\ee
One can verify directly that $\langle \omega x^\LCp\rangle = \pm \pi/2$, independent of~$r$. The instantons therefore circulate the two field extrema. For small $r$ the instantons are approximately circular again. As $r\to 1$ from below, the instantons expand up toward, but never reach, the singularity in $\mathcal{E}$ at complex infinity, see Fig.~\ref{FIG:SINUS1}. For $r=1$ the instantons `break', i.e.~fail to be periodic.

The $x^\LCm$ instantons are
\be\label{SOL:SINUS2}
	x^\LCm(\tau) = c^\LCm \mp \frac{\omega}{2\mathcal{E}^2}\big(r e^{\pm 2n\pi i \tau} - r^{-1} e^{\mp 2n\pi i \tau}\big)  \;,
\ee
also shown in Fig.~\ref{FIG:SINUS1}. (The Keldysh parameter is $\gamma=\omega/\mathcal{E}$~\cite{Dunne:2005sx}.) For $r<1$ the expectation value $\langle\mathcal{E}\rangle$ of the electric field contributing to pair production is
\be
	\langle \mathcal{E} \rangle = \mathcal{E}_0\bigg[ \pm \tau + i \frac{\log[1+r^2 e^{\pm 4\pi n i \tau}]}{2\pi n} \bigg]\bigg|_0^1 = \pm \mathcal{E}_0 \;,
\ee
confirming again that $\langle \mathcal{E}(x^\LCp) \rangle= \mathcal{E}(\langle x^\LCp\rangle)$. Hence we see that the instantons in a sinusoidal field also give local contributions, even though the field changes sign.

It is interesting to speculate on how a {\it nonlocal} contribution could arise in general. One candidate is an instanton which circulates multiple poles (zeros of the potential) and therefore yields contributions from several, rather than a single, point. Recall from (\ref{PERIODICITET}) that periodicity requires $a \langle \mathcal{E}\rangle = 2n\pi$; this implies that
\be\label{AP}
	1 = \frac{a \langle \mathcal{E}\rangle}{2n\pi}  = \frac{1}{2\pi i } \oint\limits_{\text{inst.}} \frac{\mathcal{E}}{\mathcal{A}} \;,
\ee
which is just the logarithmic derivative of $\mathcal{A}$. The argument principle then requires that the number of zeros $N$ and number of poles $P$ (counted with multiplicity) which the instanton circulates must obey
\be\label{NP}
	N - P = 1\;.
\ee
It can easily be confirmed that in the constant, Sauter and sinusoidal fields above we have $N=1$ and $P=0$, consistent with (\ref{NP}).

It follows that an instanton which circulates multiple zeros of the potential, $N>1$, must also circulate poles of the potential, $P>0$, in order to satisfy (\ref{NP}). For the sinusoidal field, this tells us that an instanton which circles {\it both} the positive and negative field maxima, i.e.~two poles of the potential, $N=2$, must also circulate a simple pole, $P=1$. However, sine is an entire function and the only pole of the field is at infinity. We might try to force one of our solutions past this pole by taking $r>1$; doing so, though, the instanton circling the e.g.~positive field maximum instead simply jumps to circulate the negative field maximum with $r\to 1/r$, as can be verified from the explicit expression (\ref{SOL:SINUS}).

So, the instantons in (\ref{E:SINUS}) localise around single field maxima, as for the Sauter field, above. One can though imagine that instantons with $P\not=0$ exist, for example, in fields with less symmetry than (\ref{E:SINUS}), and it is certainly straightforward to construct examples. The question of how or if such instantons contribute to $\Gamma$ is an interesting topic for future study.



%
%

%
\subsection{Invariance}
We turn finally to the behaviour of the instantons under changes in $r$.  The instantons (\ref{SOL:KONST}), (\ref{SOL:SAUTER}), (\ref{SOL:SINUS}), (\ref{SOL:SINUS2}) all behave differently under such changes, depending on the form of $\mathcal{E}(x^\LCp)$, but the classical action is always invariant.  We unify this behaviour as follows.

Observe that $r$ may be absorbed into the exponents in all our expressions by writing $r \exp(2\pi n i\tau)\to \exp(2\pi n i(\tau-\tau_0))$. As already stated, a real $\tau_0$ can be removed by a proper time reparameterisation without physical consequence. What we have found is that the pair production probability in fields $\mathcal{E}(x^\LCp)$ is also invariant under `generalised reparemeterisations' with $\tau_0$ \textit{imaginary}. Even though the instantons change under these reparameterisations, their classical action is invariant. The reason is simply that the complex reparameterisations correspond to contour deformations allowed by the residue theorem, which leave the value of the contour integral invariant.

\section{Conclusions}\label{Conc}
The worldline instantons which give the leading contribution to the pair production probability in electric fields $E(x^\LCp)$ are complex. The instanton contributions are contour integrals over the instantons themselves. These contributions are invariant under certain transformations which may be interpreted as complex reparameterisations (and which for constant fields resemble boosts). This complex reparameterisation invariance simply expresses the invariance of complex line integrals under contour deformation, as per Cauchy's residue theorem. 
It follows that even though the instantons are nontrivial loops, they contribute only residues from poles at certain {\it points}. This explains the locality of the effective action found previously~\cite{Tomaras:2000ag,Tomaras:2001vs}. We have shown that this structure exists even for the well-studied case of a constant electric field.

The poles are located at the zeros of the potential which, from (\ref{A}), is the velocity $\dot{x}^\LCp$ analytically continued off the loop. Thus our results are consistent with the Minkowski space worldline description~\cite{Ilderton:2014mla}, in which all contributing loops must obey $\dot{x}^\LCp\equiv 0$ (illustrating the nontrivial role of lightfront zero modes in pair production, see~\cite{Ji,ZM,Brodsky:1997de,Heinzl:2000ht,Tomaras:2000ag,Tomaras:2001vs,Ilderton:2014mla} and references therein).

To what extent can this elegant view of pair production be extended to e.g.~time dependent electric fields $E(t)$, in which instantons may also be complex and where interference terms are important~\cite{Dumlu:2011cc}? If the instantons contribute as contour integrals then they must circulate a structure other than a single pole, since the locally constant approximation does not hold in fields $E(t)$. Given that solutions of the equations of motion in that case involve a square root, and given the quantum mechanical complex instantons of~\cite{Kim:2007pm}, it seems that branch cuts will be relevant. This is confirmed in~\cite{US2}.

Finally, we hope that the structure we have uncovered here will be helpful when considering the important and challenging problem of pair production in electromagnetic fields with multi-dimensional inhomogeneities.\\[1pt]

\acknowledgements
A.I.~and G.T.~are supported by the Swedish Research Council, contract 2011-4221.

\appendix

\end{document}